\documentclass[conference]{IEEEtran}

\ifCLASSINFOpdf
\else
\fi
\usepackage{amsfonts}
\usepackage{amssymb}
\usepackage{amsmath}
\usepackage{graphicx}
\usepackage{subfigure}
\usepackage{algorithm}
\usepackage{algpseudocode}
\usepackage{cite}
\usepackage{multirow}
\usepackage{array}
\usepackage{multirow}
\usepackage{array}
\usepackage{soul}
\usepackage{slashbox}
\usepackage{algorithm}
\usepackage{algpseudocode}

\usepackage{booktabs}
\usepackage{threeparttable}

\usepackage{stfloats}

\begin{document}

\title{Divisible Load Scheduling in Mobile Grid based on Stackelberg Pricing Game}
\author{\authorblockN{Jiadi~Chen, Qiang~Zheng, Hang~Long, Wenbo~Wang}
\authorblockA{Wireless Signal Processing and Network Lab, Key Laboratory of Universal Wireless Communication, \\
Ministry of Education, Beijing University of Posts and Telecommunications, Beijing 100876, China\\
Email: chenjd@bupt.edu.cn}} \maketitle

\maketitle

\begin{abstract}

Nowadays, it has become feasible to use mobile nodes as contributing entities in computing systems. In this paper, we consider a computational grid in which the mobile devices can share their idle resources to realize parallel processing. The overall computing task can be arbitrarily partitioned into multiple subtasks to be distributed to mobile resource providers (RPs). In this process, the computation load scheduling problem is highlighted. Based on the optimization objective, i.e., minimizing the task makespan, a buyer-seller model in which the task sponsor can inspire the SPs to share their computing resources by paying certain profits, is proposed. The Stackelberg Pricing Game (SPG) is employed to obtain the optimal price and shared resource amount of each SP. Finally, we evaluate the performance of the proposed algorithm by system simulation and the results indicate that the SPG-based load scheduling algorithm can significantly improve the time gain in mobile grid systems.

\end{abstract}
\footnotetext[1]{ This work is supported in part by the Fundamental Research Funds for the Central Universities (No.2014ZD03-02), National Key Technology R\&D Program of China (2014ZX03003011-004) and China Natural Science Funding (61331009).  }

\section{Introduction}

Computational grids and clusters have been widely used to solve computationally-intensive problems~\cite{grid}. With rapid developments in computer science and manufacturing technologies, many mobile devices, e.g., smart phones, laptop computers, intelligent robots, etc., have growing processing abilities and storage capacities. Together with the advances in wireless communication,  it has become feasible to use mobile nodes as contributing entities to grid systems. In another words, mobile devices can act as cloud resources under some circumstances~\cite{MCC}. 

In this paper, we consider the systems in which mobile nodes can share computing resources with each other through wireless connections.  The resource consumer (RC) distribute computational loads to multiple mobile resource providers (RPs) to realize parallel processing. During this process, the load scheduling problem is highlighted.
We partition a divisible computing task~\cite{divisible} into multiple subtasks, which have different computing volumes and different input/output data volumes. The load scheduling problem is to determine the partition policy of the overall computational task, then distribute each subtask to the corresponding RP.

Many approaches have been proposed to integrate mobile nodes with grid computing systems,  in which the grid schedulers can be categorized by the optimization objectives. Authors in~\cite{makespan1} --\cite{makespan3} try to optimize the makespan for some time sensitive tasks. In~\cite{throughput1} --\cite{throughput3}, the performance metric is the throughput, which is the key factor to determine the QoS of some certain services, e.g., the data streaming service~\cite{datastream}. There are other approaches that aims at improving the energy efficiency of mobile devices such as~\cite{energy}.

In this paper, we try to minimize the makespan of the overall task by introducing a novel computational load scheduling algorithm into mobile grid systems. Main contributions are summarized as follows:
\begin{itemize}
  \item [i)] The reward for sharing idle resources is considered in the problem formulation. Therefore, a buyer-seller model in which the task sponsor can inspire other mobile devices to share their computing resources by paying certain profits, is proposed. The Stackelberg Pricing Game (SPG)~\cite{SPG} is employed to obtain the optimal price and shared amount of each SP's resources.
  \item [ii)] Two typical computing scenarios are analyzed, i.e., the proxy-based mobile grid and the mobile ad hoc grid. The RC can be a mobile device or a computing entity in the wired domain, and the RPs can access to a base station/access point or be in a self-organizing way. These two scenarios are highly applicable in the real world systems.
  \item [iii)] Simulation results indicate that the proposed SPG-based scheduling algorithm can significantly improve the time gain in mobile grid systems and is proved with good convergency.
\end{itemize}

The remainder of this paper is organized as follows. The system model is described in Section II. In Section III, the problem formulation and solution is discussed. In Section IV, the simulation results are presented and analyzed. Finally the conclusion is drawn in Section V.

\section{System Model}

\subsection{Computing Scenario}

\begin{figure}
\centering
\includegraphics[width=3.3in]{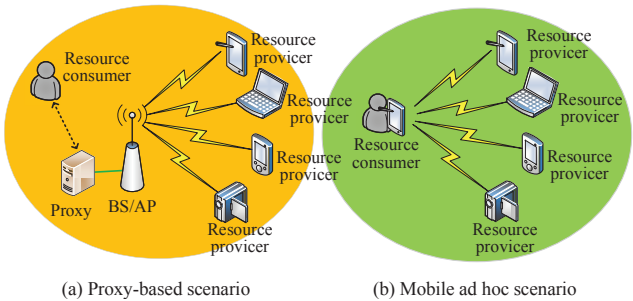}
\vspace{-10pt}
\caption{Computing scenarios.}
\vspace{-10pt}
\label{application_scenario}
\end{figure}

As Fig. 1 shows, we consider two computing scenarios in this paper, i.e., the proxy-based mobile grid and the mobile ad hoc grid. In both scenarios, the grid system consists of a variety of mobile devices such as smart phones, laptops, etc., which can act as the resource providers (RPs). The proxy-based mobile grid has one base station (BS) such as a femtocell~\cite{femto}, through which the mobile devices can communicate with each other.  While in the mobile ad hoc grid, the RC is a mobile device which distributes the computation load and collect the computation results through device-to-device (D2D) links.

These two scenarios are widely applicable in the real world, based on which the load scheduling problem for mobile grid systems is formulated. Assume that there are $K$ RPs in the mobile grid and each RP has a certain amount of computing resources, e.g., CPU slices. These computing resources can be divided into two parts, i.e., the dedicated computing resource (DCR) and the sharable computing resource (SCR). The former is dedicated for each device to implement basic computing while the latter can be shared for grid computing.

Assume that the RC has certain computing task to perform and it may want to borrow the RPs' SCRs for cooperation. We make the following assumptions:

i) The overall task can be arbitrarily partitioned into multiple subtasks, i.e., computation loads, for multiple RPs to conduct parallel processing.
ii) Each subtask has a certain volume of input data which needs to be transmitted to the corresponding RP. There exists a line relationship between the computing volume and input data volume for each subtask;
iii) The computing capacity for unit amount of SCR is fixed and the processing time for each subtask is inversely proportional with the occupied SCR amount;
iv) The data volume of computing result of each subtask is relatively small so that the result returning delay can be ignored in the problem formulation.

\subsection{System Model}

\subsubsection{Single-RP scenario}

We first consider the simplest case in which there are only one RC $c$ and one RP $v$. Denote the input data volume and the computing volume of the overall task as $S$ and $V$, respectively. The available SCR amount of the RC and RP to process the task are denoted as $C_c$ and $C_v$. According to assumption (iii) we have $S = \gamma V$, where $\gamma$ is the balance factor. If the RC perform the overall task by itself, the makespan is ${T_l} = {V \mathord{\left/{\vphantom {V {{C_c}}}} \right.\kern-\nulldelimiterspace} {{C_c}}}$.

In the parallel processing scenario, we assume that the channel state information (CSI) can be perfectly known by both RC and RP~\cite{channel}. The maximum achievable data rate of the wireless link between RC (BS/AP) and RP can be given by
\begin{equation}\label{eq1}
\begin{array}{l}
{R_{c,v}} = \log \left( {1 + \rm{SNR}} \right) = \log \left( {1 + \dfrac{{{p_c}\left| {{h_{c,v}}} \right|}}{{{\sigma ^2}}}} \right)
\end{array}
\end{equation}
where $p_c$ is the transmission power (in a proxy-based mobile grid, $p_c$ is the power of BS/AP, in the mobile ad hoc grid, $p_c$ is the power of the RC device), which is a constant value since the power control is not considered in this paper. $\rm{SNR}$ stands for the signal-to-noise ratio. ${{h_{c,v}\sim \mathcal{CN}\left( {0,1} \right)}}$ represents the channel fading coefficient of the link and ${{\sigma ^2}}$ is the additive white noise power at the receiver.\par

Let $\beta ,\,0 \le \beta  \le 1$ denote the proportion of computation load allocated to the RP. To minimize the makespan, the parameter $\beta$ should satisfy
\begin{equation}\label{eq2}
\frac{{\left( {1 - \beta } \right)V}}{{{C_c}}} = \frac{{\beta S}}{{{R_{c,v}}}} + \frac{{\beta V}}{{{C_v}}}.
\end{equation}
By solving (\ref{eq2}), we get
\begin{equation}\label{eq3}
{\beta ^ * } = \frac{{{R_{c,v}}{C_v}V}}{{S{C_v}{C_c} + {R_{c,v}}{C_c}V + {R_{c,v}}{C_v}V}}.
\end{equation}
In this case, the task makespan can be expressed as
\begin{equation}\label{eq4}
 \begin{array}{l}
{T_c} = {T_t} + {T_p}= \dfrac{{{\beta ^ * }\gamma V}}{{{R_{c,v}}}} + \dfrac{{{\beta ^ * }V}}{{{C_v}}},
\end{array}
\end{equation}
where $T_t$ denotes the wireless transmission time and $T_p$ is the task processing time. Therefore, the time saved by introducing the parallel processing can be expressed as
\begin{equation}\label{eq5}
\begin{array}{l}
{T_{sv}} = \max \left( {{T_l} - {T_c},0} \right)= \max \left( {\dfrac{V}{{{C_c}}} - \dfrac{{{\beta ^ * }\gamma V}}{{{R_{c,v}}}} - \dfrac{{{\beta ^ * }V}}{{{C_v}}},0} \right).
\end{array}
\end{equation}
Obviously, $T_{sv}$ is an increasing function of the shared SCR amount $C_v$. However, the $C_v$ is finite and the RP is also stingy to offer the SCRs unless there are satisfactory returns. So the key issue is how to allocate the computation load and set a reasonable price to meet the optimization objective, i.e., minimizing the makespan of the overall task.
\subsubsection{Multi-RP Scenario}
In this part, we extend the single-RP scenario to multi-RP Scenario. Assume that in the mobile grid there are multiple available RPs, the set of which is denoted as $\mathcal {A}_c$. The maximum data rate that the $j$-th RP can achieve is expressed as
\begin{equation}\label{eq6}
{R_{c,j}} = \sum\limits_{k = 1}^{{N_{\rm{RB}}}} {{s_{k,j}}\log \left( {1 + \frac{{{p_c}\left| {{h_{c,j}}} \right|}}{{{\sigma_j ^2}}}} \right),\;j \in {\mathcal {A}_c}},
\end{equation}
where $h_{c,j}$ represents the channel gains from RC (or BS/AP) to the $j$-th RP and ${{\sigma_j ^2}}$ is the additive white noise power at the $j$-th RP's receiver. $N_{\rm{RB}}$ denotes the number of transmission resource blocks in the system. In a proxy-based mobile grid, transmission resource blocks can be frequency resource blocks (in FDD system) or time slots (in TDD system). While in a mobile ad hoc grid, the transmission resources blocks can only be time slots for the FDMA is not supported in the D2D communication scene. ${s_{k,j}} \in \{ 0,1\} $ denotes the allocation indicator for the $j$-th RP at the $k$-th resource block. $s_{k,j}=1$ ($s_{k,j}=0$) indicates that the $k$-th resource block is (not) allocated to the $j$-th RP.

Let ${\beta _j},\,j \in {\mathcal {A}_c}$ represent the proportion of the overall computation load that be allocated to the $j$-th RP. $\beta _0$ is the computation load that remains to be processed by RC itself and there has the constraint ${\beta _0} + \sum\nolimits_{j \in {{\cal A}_c}} {{\beta _j}}  = 1$.

Obviously, $\beta_j$ must fulfill the same constrain as (\ref{eq2}) to minimize the makespan, i.e.,
\begin{equation}\label{eq7}
\frac{{{\beta _0}V}}{{{C_c}}} = \frac{{{\beta _j}S}}{{{R_{c,j}}}} + \frac{{{\beta _j}V}}{{{C_j}}},\;\;j \in {\mathcal {A}_c},\;\;0 \le {C_j} \le {\overline C_j},
\end{equation}
where $C_j$ is the SCR amount the $j$-th RP agreed to share and ${\overline C_j}$ is the maximum SCR amount that the $j$-th RP can provide.\par

According to the normalization condition ${\beta _0} + \sum\nolimits_{j \in {{\cal A}_c}} {{\beta _j}}  = 1$ and (\ref{eq7}), we can achieve
\begin{equation}\label{eq8}
{\beta _j} = \frac{{{\beta _0}V{R_{c,j}}{C_j}}}{{S{C_c}{C_j} + V{C_c}{R_{c,j}}}},
\end{equation}
\vspace{-10pt}
\begin{equation}\label{eq9}
{\beta _0} = \frac{1}{{1 + \sum\nolimits_{j \in {{\cal A}_c}} {{{V{R_{c,j}}{C_j}} \mathord{\left/
 {\vphantom {{V{R_{c,j}}{C_j}} {\left( {S{C_c}{C_j} + V{C_c}{R_{c,j}}} \right)}}} \right.
 \kern-\nulldelimiterspace} {\left( {S{C_c}{C_j} + V{C_c}{R_{c,j}}} \right)}}} }}.
\end{equation}
Therefore, the time saved in multi-RP scenario can be expressed as
\vspace{-10pt}
\begin{equation}\label{eq10}
\begin{array}{l}
{T_{mv}} = \max \left( {{T_l} - {T_c},0} \right) = \max \left( {\dfrac{V}{{{C_c}}} - \dfrac{{{\beta _0}V}}{{{C_c}}},0} \right).
\end{array}
\vspace{-10pt}
\end{equation}
Similar with (\ref{eq5}), the time gain is an increasing function of the occupied SCR amount. The RC makes decision on how much SCRs to ``buy" from each SPs based on their ``price". According to assumption (iii), when the shared SCR amount of a certain SP is determined, the computation load distribute to this SP is determined, too. Therefore, the computation load scheduling problem can be solved by determining the price and sold amount of each SPs' SCRs. In next section, a Stackelberg Pricing Game (SPG) is formulated to highlight this problem and the pricing process is described.
\vspace{-0pt}
\section{Problem Formulation}
\vspace{-0pt}

\subsection{Stackelberg Pricing Game}

The SPG can be viewed as a seller-buyer interaction game. The sellers, i.e., RPs, wants to maximize their profits via setting a high price. However, the buyer, i.e., RC, attempts to maximize his own utility by lower the price. In a mobile grid, the RC can inspire RPs to share their available SCRs by paying corresponding benefits.
Thus, a SPG can be formulated with the RC and RPs as players, and the Stackelberg equilibrium can be considered as the solution of this game.
The SPG is a two-level game which can be described from two sides, i.e.,

\subsubsection{The Buyer (RC) Side}
The RC wants to acquire the most benefits, i.e., minimize the task makespan, with the least possible cost. As a result, there exists a unique demand/supply-based profile for the RC, given by solving the following optimization problem:
\begin{equation}\label{eq11}
\max \quad {U_c} = {T_{mv}} - \sum\limits_{j \in {{\cal A}_c}} {{\lambda _j}{C_j}} ,\quad s.t.\quad 0 \le {C_j} \le {\overline C_j},
\end{equation}
where ${\lambda _j}{C_j}$ is the cost paid to the $j$-th RP, $\lambda _j$ denotes the price set by the $j$-th RP for sharing one unit of SCR, and $C_j$ represents the SCR amount that the RC has ``bought'' from the $j$-th RP.
\subsubsection{The Seller (RP) Side}
The RPs charge the RC for using their SCRs and they attempt to achieve as much profits as possible. Therefore, the SPG at the RPs' side can be defined as
\begin{equation}\label{eq13}
\max \quad {U_j} = {\lambda _j}C_j^{{b_j}} - {\eta _j}C_j^{{b_j}},
\end{equation}
where ${b_j} \ge 1$ is a constant tradoff balancer.

Note that the information needs to be exchanged between the RC and RPs are the price and sold amount of each RP's SCRs.  In the following part, the properties and solution of the Stackelberg equilibrium is investigated.

\subsection{Stackelberg Equilibrium}

A typical Stackelberg game theoretic equilibrium, in which one player acts as leader and the others as followers, is that the leader sets strategy taking account of the follower's optimal response. The Stackelberg game will finally converge to the Stackelberg Equilibrium. In this subsection, we give the definition of the Stackelberg Equilibrium and analyze it in a mathematical way.\par
\emph{Definition 1:} if the parameters ${\lambda _j^{\rm{SE}}}$ and $C_j^{\rm{SE}}$ are the Stackelberg Equilibrium of the proposed SPG, the following conditions should be satisfied:
when ${\lambda _j^{\rm{SE}}}$ is fixed,
\begin{equation}\label{eq14}
{U_c}\left( {\left\{ {C_j^{\rm{SE}}} \right\}} \right) = \mathop {\sup }\limits_{0 \le {C_j} \le {{\bar C}_j}} {U_c}\left( {\left\{ {{C_j}} \right\}} \right),\quad j \in {\mathcal {A}_c},
\end{equation}
and when $C_j^{\rm{SE}}$ is fixed,
\begin{equation}\label{eq15}
{U_j}\left( {\left\{ {\lambda _j^{\rm{SE}}} \right\}} \right) = \mathop {\sup }\limits_{{\lambda _i}} {U_j}\left( {\left\{ {{\lambda _i}} \right\}} \right),\quad j \in {\mathcal {A}_c}.
\end{equation}
Next we analyze the Stackelberg Equilibrium of our proposed SPG from two aspects, i.e.,

\subsubsection{The buyer's (RC's) aspect}

As the buyer, the RC is aware of the fact that the sellers, i.e., the RPs, will choose their best response to its strategy. From the previous analysis, the RC maximizes its own utility via buying the optimal volume of SRCs based on the best responses of the followers.

Based on \textit{Definition 1}, differentiate $U_c$ with respect to $C_j$ when ${\lambda _j^{\rm{SE}}}$ is fixed, we can obtain
\begin{equation}\label{eq16}\small{
\begin{array}{*{20}{l}}
{\frac{{\partial {U_c}}}{{\partial {C_j}}} =  - \frac{V}{{{C_c}}}\frac{{\partial {\beta _0}}}{{\partial {C_j}}} - {\lambda _j}}\\
{\quad \quad  = \frac{V}{{{C_c}{{\left( {1 + \sum\limits_{j \in {{\cal A}_c}} {\frac{{V{R_{c,j}}{C_j}}}{{S{C_c}{C_j} + V{C_c}{R_{c,j}}}}} } \right)}^2}}} \cdot \frac{{{V^2}{C_c}R_{c,j}^2}}{{{{\left( {S{C_c}{C_j} + V{C_c}{R_{c,j}}} \right)}^2}}} - {\lambda _j}}\\
{\quad \quad  = \frac{{\beta _0^2{V^3}R_{c,j}^2}}{{{{\left( {S{C_c}{C_j} + V{C_c}{R_{c,j}}} \right)}^2}}} - {\lambda _j}}
\end{array}}
\vspace{-15pt}
\end{equation}
and
\begin{equation}\label{eq17}%\small{
\begin{array}{l}
\quad \frac{{{\partial ^2}{U_c}}}{{\partial C_j^2}} = \frac{{2{\beta _0}\frac{{\partial {\beta _0}}}{{\partial {C_j}}}{V^3}R_{c,j}^2\left( {S{C_c}{C_j} + V{C_c}{R_{c,j}}} \right) - 2\beta _0^2{V^3}R_{c,j}^2}}{{{{\left( {S{C_c}{C_j} + V{C_c}{R_{c,j}}} \right)}^3}}} < 0
\end{array}.%}
\end{equation}
It can be obviously concluded  from (\ref{eq16}) and (\ref{eq17}) that the utility function is concave. Therefore, by setting (\ref{eq16}) to zero we can derive the optimal solution $C_j^ * $ as
\begin{equation}\label{eq18}
C_j^* = {u_j}\sqrt {{1 \mathord{\left/
 {\vphantom {1 {{\lambda _j}}}} \right.
 \kern-\nulldelimiterspace} {{\lambda _j}}}}  - {v_j},
 \end{equation}
where
\begin{equation}\label{eq181}
\begin{array}{l}
\small{{u_j} = \dfrac{{V{R_{c,j}}\sqrt V }}{{S{C_c} + V{R_{c,j}} + S{C_c}{w_j}}}}\\
\small{{v_j} = \dfrac{{V{C_c}{R_{c,j}}\left( {1 + {w_j}} \right)}}{{S{C_c} + V{R_{c,j}} + S{C_c}{w_j}}}}\\
\small{{w_j} = \sum\limits_{i \in {\mathcal {A}_c},i \ne j} {\dfrac{{V{R_{c,i}}{C_i}}}{{S{C_c}{C_i} + V{C_c}{R_{c,i}}}}}}
\end{array}.
\end{equation}

Finally, due to the boundary conditions in (\ref{eq11}), the optimal amount of SCRs bought from the $j$-th RP can be expressed as
\begin{equation}\label{eq19}
C_j^ *  = \min \left( {C_j^ * ,{{\overline C}_j}} \right)
\end{equation}

\subsubsection{The Seller's (RPs') aspect}

The RPs' goal is to maximize their profits via setting the optimal unit price for the SCR. By differentiating the RPs' utility functions with respect to $\lambda_j$ and setting it to zero, we get
\begin{equation}\label{eq20}\small{
\frac{{\partial {U_j}}}{{\partial {\lambda _j}}} = {\left( {C_j^*} \right)^{{b_j}}} + {\lambda _j}{b_j}{\left( {C_j^*} \right)^{{b_j} - 1}}\frac{{\partial C_j^*}}{{\partial {\lambda _j}}} - {\eta _j}{b_j}{\left( {C_j^*} \right)^{{b_j} - 1}}\frac{{\partial C_j^*}}{{\partial {\lambda _j}}} = 0.}
\end{equation}
By deriving the above equation, we obtain
\begin{equation}\label{eq21}
{\left( {C_j^*} \right)^{{b_j} - 1}}\left( {C_j^* + {b_j}\frac{{\partial C_j^*}}{{\partial {\lambda _j}}}\left( {{\lambda _j} - {\eta _j}} \right)} \right) = 0
\end{equation}
From (\ref{eq21}), following conclusion can be achieved: $C_j^ *  = 0$, which indicates that the consumer won't buy any SCR from the $j$-th RP.
Otherwise,
\begin{equation}\label{eq22}
C_j^* + {b_j}\frac{{\partial C_j^*}}{{\partial {\lambda _j}}}\left( {{\lambda _j} - {\eta _j}} \right) = 0,
\end{equation}
which has a unique solution, i.e.,
\begin{equation}\label{eq23}
\lambda _j^ *  = {\eta _j} - \frac{{C_j^*}}{{{b_j}{{\partial C_j^*} \mathord{\left/
 {\vphantom {{\partial C_j^*} {\lambda _j^ * }}} \right.
 \kern-\nulldelimiterspace} {\lambda _j^ * }}}}
\end{equation}

\subsection{Solution of Stackelberg Equilibrium}

In this subsection, a iteration algorithm is employed to obtain the optimal price and shared resource amount. Let ${\bf{\lambda }} = {\left\{ {{\lambda _j}} \right\}_{j \in {\mathcal {A}_c}}}$ be the price vector  and ${{\mathop{\rm F}\nolimits} _i}\left( {\bf{\lambda }} \right)$ represent the updating function, i.e.,
\begin{equation}\label{eq24}
{\lambda _j} = {{\mathop{\rm F}\nolimits} _i}\left( {\bf{\lambda }} \right) = {\eta _j} - \frac{{C_j^*}}{{{b_j}{{\partial C_j^*} \mathord{\left/
 {\vphantom {{\partial C_j^*} {\partial {\lambda _j}}}} \right.
 \kern-\nulldelimiterspace} {\partial {\lambda _j}}}}}.
\end{equation}
The updating function can be rewritten in a vector form as
\begin{equation}\label{eq25}
{\bf{\lambda }}\left( {t + 1} \right) = {\bf{F}}\left( {{\bf{\lambda }}\left( t \right)} \right),
\end{equation}
where ${\bf{F}} = {\left\{ {{F_j}} \right\}_{j \in {\mathcal {A}_c}}}$ and $t$ denotes the iteration time. The solution of the SPG can be divided into four steps, which are described in
\textbf{Algorithm 1}. %The convergence of the algorithm is proofed in the Annex.

\begin{algorithm}
\caption{ \quad The Solution of SPG} \label{alg}
\begin{algorithmic}
\State \textbf{Step 1} - \small{\emph{Initialization}}\par

\small{For RPs, initialize the price vector ${\bf{\lambda }}\left( 0 \right) = {\left\{ {{\lambda _j}\left( 0 \right)} \right\}_{j \in {\mathcal {A}_c}}}$ and inform it to the RC. For the RC, initialize its optimal amount of bought SRCs, i.e., ${C_j}\left( 0 \right)$. Set the iteration index $t=1$.}

\State \textbf{Step 2} - \emph{RC's update}\par

\small{Use ${\bf{\lambda }}\left( t \right)$ to obtain the optimal amount of SCRs for each RP according to (\ref{eq18}), get ${\bf{C}}\left( {t + 1} \right) = {\left\{ {C_j^ * \left( {t + 1} \right)} \right\}_{j \in {\mathcal {A}_c}}}$.}

\State \textbf{Step 3} - \emph{RPs' update}\par

\small{Based on ${\bf{C}}\left( {t + 1} \right)$ obtained in \textbf{Step 2}, calculated the optimal price $\lambda _j^ * $ for each RP according to (\ref{eq24}), which are denoted as ${\bf{\lambda }}\left( {t + 1} \right)$. Then update the load scheduling decision $\beta_j$ and update the iteration index $t$ by $t+1$.}\par

\State \textbf{Step 4}\par

\small{Repeat \textbf{Step 2} and \textbf{Step 3} until both ${\bf{\lambda }}\left( t \right)$ and ${\bf{C}}\left( t \right)$ do not change any longer or the differential value between two successive iteration is smaller than a predefined threshold. Take the final Stackelberg Equilibrium as the solution to the SPG.}
\end{algorithmic}
\end{algorithm}

\section{Simulation Results and Analysis}

\begin{figure*}
\centering
\subfigure[Optimal price of per unit SCR verse the iteration time.]{ \label{nr} %% label for first subfigure
\vspace{-10pt}
\includegraphics[width=0.41\textwidth]{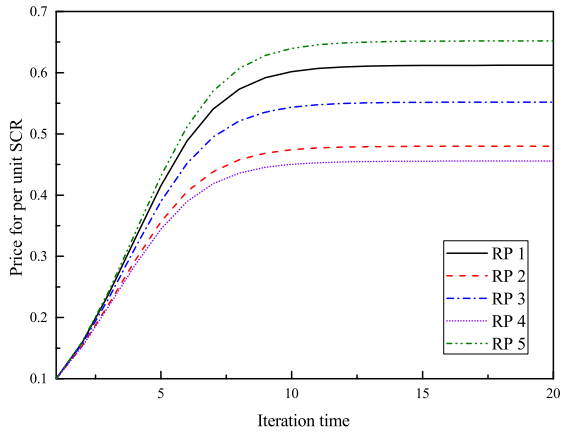}} %\hspace{-0.95in}
\subfigure[Optimal SCR amount the RC bought from each RP.]{ \label{mr} %% label for second subfigure
\vspace{-10pt}
\includegraphics[width=0.41\textwidth]{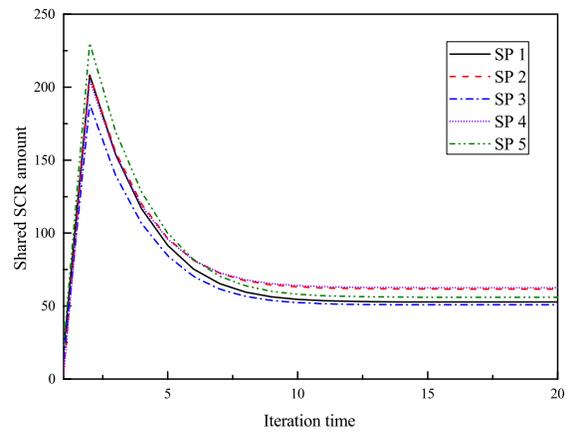}}
\vspace{-5pt}
\caption{Convergence performance of the proposed SPG algorithm versus the iteration time.} \label{drop} %% label for entire figure \end{figure}
\vspace{-15pt}
\end{figure*}

We consider the proxy-based mobile grid in the simulation and the scenario is described as follows. A number of mobile devices which act as the RPs, spread over a square room. the RC is a computation node in the wired domain. The wireless AP is located at the center of the room, by which the RC and RPs can communicate with each other. For simplicity, we suppose that the mean $\rm{SNR}$ from AP to the farthest RP (located at the room edge) equals to 0 dB. The Adaptive Coding and Modulation (AMC) and 16 Modulation and of Coding Schemes (MCSs)~\cite{3GPP} are adopted. The channel fading coefficient is i.i.d. over scheduling slots. Main parameter values in the simulation are listed in TABLE I.

\begin{table}
\centering
\vspace{-10pt}
\caption{Simulation parameters }
\vspace{-10pt}
\begin{tabular}{|m{4cm}<{\centering}|m{3cm}<{\centering}|}
\hline \textbf{Parameter} & \textbf{Value} \\
\hline \small{Room size} & \small{$10\,{\rm{m}} \times 10\,{\rm{m}}$} \\
\hline \small{SP number} & \small{3 to 12}\\
\hline \small{Distance between devices} & \small{3 to 10 m }\\
\hline \small{Pathloss model} & \small{$P\left( d \right) = P{d^{ - \alpha }}$} \\
\hline \small{$\alpha$} & \small{4} \\
\hline \small{Bandwidth} & \small{10 MHz} \\
\hline \small{Carrier frequency} & \small{2 GHz} \\
\hline \small{Scheduling slot length} & \small{1 ms} \\

\hline
\end{tabular}
\vspace{-20pt}
\label{TAB_SIMUPARA}

\end{table}

First the convergence performance of the proposed SPG algorithm is evaluated. When there are 5 RPs in the system, Fig. 2 (a) shows the changing process of each SP's price for one unit SCR versus the iteration time. Each SP's SCR price converges to one of the 5 different values with the same initial value 0.1. The higher price is dealed for higher service quality, e.g., higher transmission rate. Similarly, Fig. 2 (b) shows the convergence process of the SCR amount the RC bought from each RP. From  Fig. 3 (a) and (b), we can conclude that the proposed algorithm has a good convergence performance, i.e., converges within 15 iteration time under 5-SP scenario.

In the simulation, we evaluate the time gain by realizing the mobile grid computing. Two representative scheduling schemes, i.e., the Round Robin (RR) scheduling and the MaxWeight scheduling~\cite{MW}, are adopted to determine the allocation policy of the wireless transmission resources, i.e., wireless resource blocks. Fig. 4 illustrates the reduction ratio of the task makespan versus the available SP number, and indicates that the adopted wireless resource scheduling scheme can effect the performance of the SPG-based algorithm. Therefore, the proposed computation load scheduling algorithm should cooperate with advanced wireless resource allocation schemes to achieve further performance improvement.

Additionally, we consider another simple and straightforward computation load scheduling policy for comparision, i.e., the SC borrowing equal amount (denoted as $a$) of SCRs from each RP. We refer this policy as the ESE algorithm and $a$ is equal to the mean value of the borrowed SCR amount determined by the SPG-based algorithm under same simulation parameters. From Fig. 4 we can see that with the increase of RP number, the time gain by realizing the grid computing is also increasing, and the proposed SPG-based algorithm outperforms the ESE algorithm. When the SP number grows to 12, the ratio of saved time under the proposed algorithm can be up to 80\% compared with the makespan in the nonparallel case. Therefore, we can conclude that the proposed computation load scheduling algorithm can significantly improve the computing efficiency in mobile grid systems.

\begin{figure}
\centering
\includegraphics[width=0.41\textwidth]{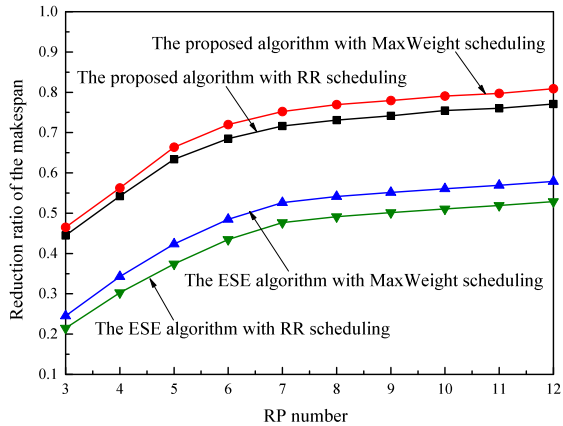}
\vspace{-10pt}
\caption{Proportion of Saving time verse Number of Volunteer.}
\vspace{-15pt}
\label{result3}
\end{figure}

\vspace{-5pt}
\section{Conclusion}

In this paper, we consider the divisible computation load scheduling problem in grid systems where the mobile devices can share their idle resources for parallel processing. Two main computing scenarios are considered, i.e., the proxy-based mobile grid and the mobile ad hoc grid. The task sponsor want to borrow as much resource as possible to reduce the task makespan, while the resource providers are not willing to share their resources unless there are satisfactory returns. Based on this, a buyer-seller model is designed and a SPG-based algorithm is proposed to obtain the optimal unit price and shared amount of the computing resources. For performance evaluation, a system simulation is conducted and the results indicate that the proposed SPG-based load scheduling algorithm can significantly improve the computing efficiency in mobile grid systems (80\% reduction on the makespan when SP number is 12) as well as a excellent convergence speed (within 15 iteration times when SP number is 5).

\end{document}